# Alternatives to the Journal Impact Factor:

# *I3* and the Top-10% (or Top-25%?) of the Most-Highly Cited Papers

Loet Leydesdorff

Amsterdam School of Communication Research (ASCoR), University of Amsterdam

Kloveniersbugwal 48, 1012 CX Amsterdam, the Netherlands;

loet@leydesdorff.net ; http://www.leydesdorff.net

**Abstract**

Journal Impact Factors (IFs) can be considered historically as the first attempt to normalize citation distributions by using averages over two years. However, it has been recognized that citation distributions vary among fields of science and that one needs to normalize for this. Furthermore, the mean—or any central-tendency statistics—is not a good representation of the citation distribution because these distributions are skewed. Important steps have been taken to solve these two problems during the last few years. First, one can normalize at the article level using the citing audience as the reference set. Second, one can use non-parametric statistics for testing the significance of differences among ratings. A proportion of most-highly cited papers (the top-10% or top-quartile) on the basis of fractional counting of the citations may provide an alternative to the current IF. This indicator is intuitively simple, allows for statistical testing, and accords with the state of the art.

# 1. Introduction

In the lead article of this topical issue entitled "Impact Factor: Outdated artefact or stepping-stone of journal certification?" Jerome K. Vanclay focuses primarily on data errors in the database of Thomson Reuters, but less on the statistics of the Impact Factor (IF) as an indicator. The author mentions that the third decimal is provided unnecessarily (in order to minimize the number of tied places; cf. Garfield, 2006) and that citation distributions are highly skewed (Seglen, 1992, 1997). However, the possible flaws introduced by using averages of these skewed distributions across the file are not elaborated, and significance of differences between impact factors or the statistical estimation of error in the measurement do not enter into the discussion.

The technical problems in the database can increasingly be corrected with further investments in the data processing, but flaws in the data analysis provide an opportunity for scientometric improvement of the indicator. The merit and quality of an indicator depends on its statistical properties and the evaluation of its validity and reliability. In this contribution, I focus on these issues: does the IF measure impact? How can one account for differences in citation behavior among fields of science? How can one appreciate the skewness in citation distributions using appropriate statistics?

## 2. Stating the problem

Using the same scales, Figure 1 shows—as an example—the distributions of the IFs-2010 of 125 journals classified in the Web of Science under "sociology" (the subject category "XA" in the database) to the left, and the 73 journals classified as "psychology" ("VI") to the right. The two means—0.870 (± 0.061) and 2.555 (± 0.321), respectively—are significantly different ($p < 0.01$).[1]

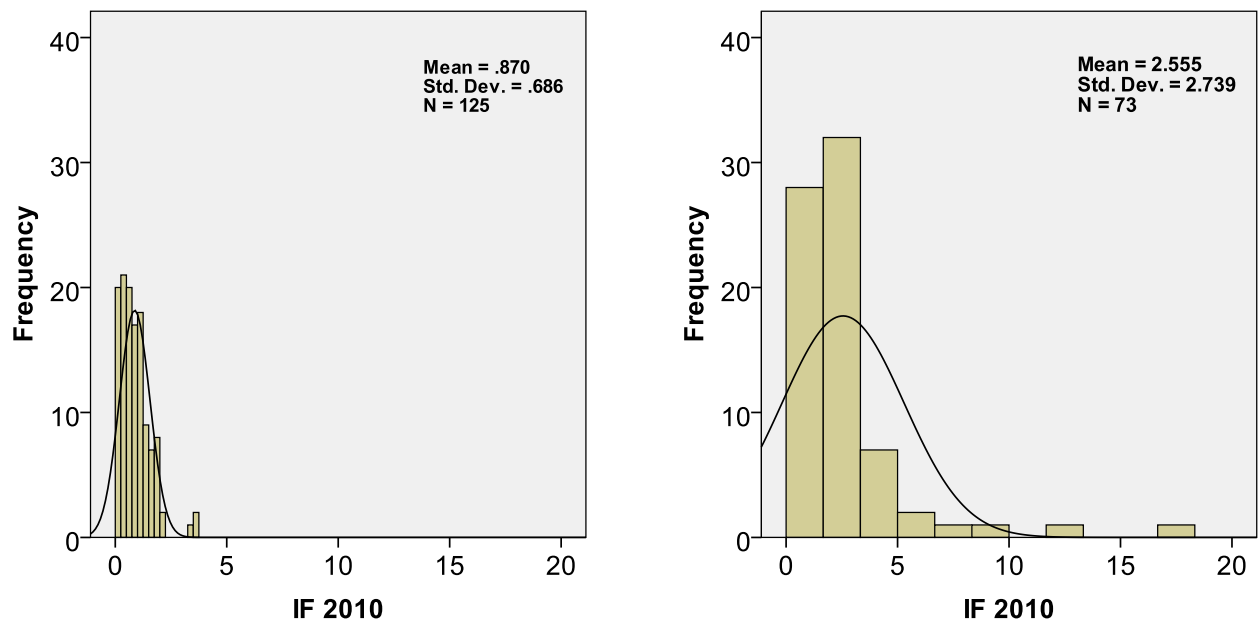

**Figure 1**: Impact factors (2010) of 125 journal in the WoS Category "sociology" (XA; left) compared with 75 journals in the WoS Category "psychology" (VI; right)

Sociology and psychology are neighbouring disciplines. The database additionally distinguishes a category "social psychology" with 56 journals (of which 4 overlap with sociology). The mean IF-2010 of this set is 1.499 (± 0.169). This distribution is again significantly different from that of sociology journals at the 1% level.[2] Social psychology can be considered as a subfield of psychology, but the distribution of the impact factors of this subfield is nevertheless significantly different from that of the psychology journals at the 5% level.

---

[1] IFs can be considered as means of citations (in the current year) per publication during the previous two years; therefore, one can expect these mean values to be normally distributed.

Thus, a first problem is that one is not allowed to compare IFs even across neighbouring fields and subfields. However, the delineations among fields can also be fuzzy (Boyack & Klavans, 2011; Leydesdorff, 2006). Scientometricians have used the ISI Subject Categories—recently renamed by Thomson Reuters the WoS Subject Categories—for the normalization, but journals can be attributed to more than a single field, and the attributions themselves are often erroneous (Boyack *et al.*, 2005; Rafols & Leydesdorff, 2009). In summary, the problem of the delineation of appropriate sets of journals for the comparison—reference sets—poses a problem that has hitherto remained unresolved.

But even if one accepts that one could compare within these sets—for pragmatic reasons and despite the noted problems—then the normalization in terms of mean values (or IFs as two-year averages) remains unfortunate. Yet, this has been standard practice. How should one then proceed? Schubert & Braun (1986) proposed comparing the mean of the observed citation rates (*MOCR*) in a sample under study (e.g., during an evaluation) with the mean citation rate in the reference set as the expectation (Mean Expected Citation Rate or *MECR*). These authors introduced the Relative Citation Rate (*RCR*) as the quotient: $RCR = MOCR/MECR$.[3] However, the division of two means provides a quotient without a standard error of the measurement (SEM). Consequently, scientometric bar charts and tables often fail to show error bars and to specify uncertainty.

---

[2] The four journals in the overlap were excluded from the comparison of the means. These journals are: *Deviant Behavior*, the *Int. J. of International Relations*, the *Kölner Zeitschrift für Soziologie und Sozialpsychologie*, and *Social Justice Research*.

This practice of dividing means was followed by the Center for Science and Technology Studies (CWTS) at Leiden University (Moed *et al.*, 1995) and more recently by the Centre for Research & Development Monitoring (ECOOM) at Leuven University (Glänzel *et al.*, 2009). The relative indicators were renamed with minor modifications as *CPP/FCSm* in Leiden ("the crown indicator") and *NMCR* in Leuven. In my opinion, the division of two means contains an error against the order of operations which prescribes first to divide and then to sum. Instead of $MOCR/MECR = \frac{(\sum_{obs} c_{obs})/n_{obs})}{(\sum_{exp} c_{exp})/n_{exp})}$ one should for arithmetical reasons have used the mean of the observed versus expected citation rates, or in formula format: $M(OCR/ECR) = (\sum_{i=1}^{n} \frac{observed_i}{expected_i})/n$, in which the expected citation rate is equal to the one derived from the reference set. Unlike *MOCR/MECR*, *M(OCR/ECR)* is a normal average with a standard deviation.

This problem was noted by Lundberg (2007), but ignored at the time. Only in 2010 and 2011 did it receive serious discussion in the *Journal of Informetrics* (Opthof & Leydesdorff, 2010; Van Raan *et al.*, 2010; Larivière & Gingras, 2011). CWTS in Leiden was responsive to the critique and changed the indicator within half a year (Waltman *et al.*, 2011), but the old normalization is still in place in other centers.

---

[3] One cannot test the *MOCR* against the *MECR* because the two distributions are not independent: the publication set is a subset of the reference set.

What does this discussion mean for the IF? Instead of first aggregating the numbers of citations in the current year to citable items in the previous two years (IF = $\frac{c_{-1} + c_{-2}}{p_{-1} + p_{-2}}$), one could normalize for citations to each of the previous two years separately, as follows: [ $(\frac{c_{-1}}{p_{-1}} + \frac{c_{-2}}{p_{-2}})/2$ ]. The IF would then be a moving average with a period of two (Rousseau & Leydesdorff, 2011). The difference may be marginal in most cases, but in 2009 the IFs of 8.6% of the journals would be changed in the *first* decimal! At the extremes, *Psychological Inquiry* would go from an IF-2009 of 4.050 to 9.750 and the *Annual Review of Biophysics* from 19.304 to 9.625.[4] Obviously, statistical decisions matter for the ranking: one can expect the mean of a skewed distribution to be highly sensitive to relatively minor changes in the computation.

In summary, in addition to correcting the technical errors in the database as summarized by Vanclay (this issue; cf. Leydesdorff, 2008, Table 4 at p. 285) and the arithmetic error in the calculation of these indicators, two scientometric problems remain: (*i*) how to compare "like with like" (Martin & Irvine, 1983; cf. Rafols *et al.*, in press) when the units for the comparison are so different and the differentiations not crisp, and (*ii*) how to avoid using averages over skewed distributions? In my opinion, important steps towards solutions to these problems have been taken during the last two years.

---

[4] In terms of relative decline, the journal *Oceanological and Hydrobiological Studies* would suffer most with a drop of the IF-2009 from 0.622 to 0.041.

## 3. Comparisons across fields of science

Since the 1980s scientometricians have tried to use the grand matrix of aggregated journal-journal citations for the delineation of fields of science (Doreian & Fararo, 1985; Leydesdorff, 1986; Tijssen *et al.*, 1987). This matrix can be constructed from the data in the *Journal Citation Reports* (*JCR*) which have been available (in print) since the mid-1970s. However, the emphasis remained initially on the creation of local journal maps because the decomposition of such a large file (of several thousands of journals) was computationally too intensive for the technology at the time. With the advent of Windows-95 and Pajek in 1996 the decomposition and visualization of large (citation) networks became feasible (Boyack *et al.*, 2005; Leydesdorff, 2004). The *JCRs* are electronically available since 1994.

The conclusion from this research program, in my opinion, has been that any decomposition is beset with error because the sets are not always sufficiently crisp (Leydesdorff, 2006). Furthermore, journals themselves are not homogeneous units of analysis in terms of their cognitive contents nor in terms of document types. Letters, for example, have citation half-life times completely different from review articles (Leydesdorff, 2008, p. 280). One cannot lump citable items together, and it seems that journal cannot be classified without ambiguity. Classification reduces the data into a tree-like hierarchy, whereas developments take place heterarchically. New entrants (journals), for example, may change the network thoroughly in both cognitively and policy-relevant ways (Leydesdorff *et al.*, 1994).

A solution may be to disaggregate at the level of documents. Documents can be cited in different disciplines and by different types of documents. For example, one can expect papers in the 73 psychology journals used in Figure 1 to be cited more frequently than papers in the 125 sociology journals. These differences in "citation potentials" (Garfield, 1979) can be corrected by "source-normalization" (Moed, 2010): the source of the difference is an underlying difference in the citation behavior of the citing authors. More references are expected in some fields than in others. Accordingly, each citation can be fractionally counted, that is, as one over the total number of references (1/NRef) in the citing paper. The field "NRef" is conveniently contained in the WoS database.

|  | *Annals of Mathematics* | *Molecular Cells* |
|---|---|---|
| IF 2007 | 2.739 | 13.156 |
| Fractionally counted quasi-IF-2007 on the basis of 3 years in **Scopus** [a] | 0.257 | 0.386 |
| SNIP 2007 | 4.979 | 3.696 |
| IF 2008 | 3.447 | 12.902 |
| Factionally counted quasi-IF-2008 in the **Web of Science** [b] | 1.416 | 1.143 |

**Table 1**: Comparisons between *Annals of Mathematics* and *Molecular Cells* in Scopus and WoS. Sources: [a] Leydesdorff & Opthof (2010, p. 2367) and [b] Leydesdorff & Bornmann (2011a, p. 222).

Leydesdorff & Opthof (2010) showed that this correction normalizes the huge differences in citation potentials between journals in mathematics to the extent that a leading journal in mathematics (*Annals of Mathematics*) can be ranked even more highly than a major journal in molecular biology despite the latter's (approximately) four times higher IF (Table 1). Leydesdorff & Bornmann (2011a) have computed quasi-IFs-2008 for 5742 journals in the *Science Citation Index* (available at

http://www.leydesdorff.net/weighted_if/weighted_if.xls). Using regression analysis, they showed that 81% of the variance between 13 fields of science as distinguished in the *Science and Engineering Indicators* of the US National Science Board (2010) is thus corrected, and the remaining differences among these fields are statistically non-significant (cf. Radicchi & Castellano, 2012).

In addition to this methodological advantage, a conceptual advantage of using citing papers as the reference set for normalization is the delineation *ex post* in terms of relevant audiences (Zitt & Small, 2008). Classifying an evolving system in terms of *ex ante* categories can be expected to lead to error because the classification system is then defined historically while the dynamics of science is evolutionary (Leydesdorff, 1998; Rafols *et al.*, in press). Using the metaphor of a research front (Garfield, 1972, 1979; Price, 1965), one would expect important contributions to be made also at the edges of and in between fields (Abbasi *et al.*, in press; Leydesdorff *et al.*, 1994). Authors can be cited in fields unintentionally because the intellectual organization of the sciences is self-organizing as scholarly discourses at the supra-individual and supra-institutional levels (Leydesdorff & Amsterdamska, 1991; Fujigaki, 1998).

In an evaluation of the different departments of the Tsinghua University in Beijing, for example, Zhou & Leydesdorff (2011) have shown that fractional counting can correct significantly for disadvantages of departments such as those in the arts and humanities when using scientometric evaluations. The Department of Chinese Language and Literature that has previously been rated at the 19[th] position among 27 departments, was ranked 2[nd] after the correction for citation potentials reflecting differences in citation behavior among fields of scholarly discourse.

## 4. Skewed distributions and non-parametric statistics

In the case of skewed citation distributions, one should avoid central tendency statistics, but use non-parametric statistics such as percentiles (deciles, quartiles, etc.). Bornmann & Mutz (2010) intervened in the discussion about dividing averages or averaging rates by elaborating on the metrics for the six percentile ranks used in the *Science & Engineering Indicators*: top-1%, top-5%, top-10%, top-25%, top-50%, and bottom-50% (National Science Board, 2010). Leydesdorff *et al*. (2011) elaborated these statistics, and Leydesdorff & Bornmann (2011b) applied a newly defined "Integrated Impact Indicator" (*I3*) to two groups of journals: the set of 65 journals classified in the WoS as Information & Library Science, and the 48 "multidisciplinary" journals, including journals such as *Science, Nature,* and *PNAS*.

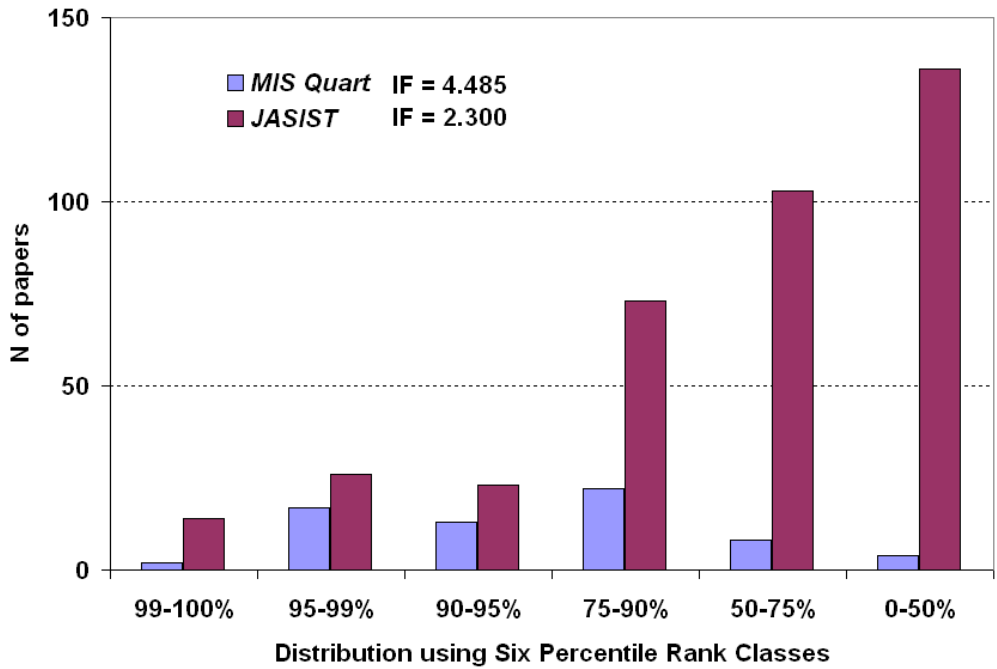

**Figure 2**: Six percentile rank classes and impact factors for *JASIST* and *MIS Quarterly* in 2009. (Source: Leydesdorff & Bornmann, 2011b, at p. 2135).

Figure 2 shows the problem. Using the reference set of 65 Information and Library Science journals, *JASIST* has higher values in *all* six classes, but its IF-2009 was only half the size of that of *MIS Quarterly*. Not only is the tale of less-frequently-cited papers in *JASIST* much larger (*N* of publications = 375), but the 66 most-cited papers in *JASIST* 2007 and 2008 are also significantly more cited than the 66 papers in the denominator of the IF-2009 of *MIS Quarterly*. Thus, the IFs erroneously give the impression that *MIS Quarterly* has an impact higher than *JASIST* (or *Scientometrics* in this set) although in fact it does not.

The misunderstanding is generated by the semantics: the words "impact" and "impact factor" or average impact have been used without sufficient distinction. An average value is determined not only by the numerators, but also the denominators. When less-cited papers are added to a set of highly-cited papers *ceteris paribus*, the total impact of these papers can be expected to increase, but the average impact may decrease. For example, the team of a leading scientist (including postdocs and Ph.D. students) will have more impact than a scientist working alone, but the team's *average* impact is lower.

Accordingly, Bensman (1996) could show that Total Citations—the numerator of the IF—could be validated by faculty significantly more than "impact factors" in the evaluation of journals. However, "total citations" are a crude measure. When publications are qualified in terms of percentiles of the citation distribution—instead of by averaging—an Integrated Impact Indicator (*I3*) can be defined as follows:

$$I3 = \sum_i x_i * n(x_i) \qquad (1)$$

In this formula, $x_i$ denotes the percentile (rank) value *i*, and *n* the number of papers with this value. Instead of averaging, the citation curves are thus integrated after proper normalization to the same scales of hundred percentiles.[5] The scaling makes the distributions comparable. One can also aggregate the percentile values into a normative evaluation scheme such the six classes used in the *US Science and Enginieering Indicators* (see Figure 2).

---

[5] The hundred percentiles can be considered as a continuous random variable or more precisely as "quantiles".

**Table 2**: *MIS Quarterly*, *JASIST,* and *Scientometrics* compared in terms of total citations, IF-2009, %I3, and %PR6 within the reference set of 65 LIS journals. (Source: Leydesdorff & Bornmann, 2011b, at p. 2139.)

| Journal | N of papers (a) | Total citations (d) | IF 2009 (c) | % I3 (b) | % PR6 (six ranks) (e) |
|---|---|---|---|---|---|
| *MIS Quart* | 66 | 847 | 4.485 [1] | 2.61 [7][+] | 2.34 [7][+] |
| *J Am Soc Inf Sci Technol* | 375 | 1975 | 2.300 [7] | 9.73 [1][+] | 8.63 [1][+] |
| *Scientometrics* | 258 | 1336 | 2.167 [10] | 7.24 [2][+] | 6.37 [2][+] |

Note. [+]$p < 0.01$; above the expectation. Ranks are added between brackets.

In Table 2, *MIS Quarterly* and *JASIST* are compared in terms of their IFs 2009 and the new indicators, and *Scientometrics* is added to the comparison for the purpose of this discussion. The first and seventh place are precisely reversed between *MIS Quarterly* and *JASIST*, and *Scientometrics* moves from the 10th place in the ranking of IFs 2009 to the second place using both *I3* (on the basis of quantiles) and the six percentile ranks (*6PR*) as indicators. All three journals are cited above expectation given the reference group of 65 LIS journals ($p < 0.01$).

The Integrated Impact Indicator (*I3*) thus combines a number of advantages:

1. *I3* values can be tested against expectations using the *z*-test for two independent proportions.
2. *I3* values are determined at the article level. A journal can thus be defined as one possible set of papers, but other aggregations remain equally possible. For example, one can also compare countries or institutions in terms of their *I3* values (Leydesdorff, 2011).
3. Percentiles can be aggregated differently in terms of the (normative) evaluation scheme chosen in a given policy context. I mentioned the evaluation scheme of six ranks used by the US-NSF, but the Research Assessment Exercises (RAE) in the UK, for example, has hitherto used 4+ classes (e.g., Rafols *et al.*, in press).

4. Rousseau (in press) noted that the popular indicator of the top-10% most-highly-cited papers (Tijssen *et al.*, 2003)—e.g., the *Excellence Indicator* of the new edition of the SCImago Institutions Rankings, and also used for the Leiden Ranking 2011/2012 (CWTS, 2011)—can be considered as a special case of two percentile rank classes. Bornmann *et al.* (in press) elaborated the test statistics for this special case.

5. *I3* values can be used across databases; for example, the user may wish to include "grey literature" or so-called non-source references (in the WoS) in the reference set (e.g., Bensman & Leydesdorff, 2009). However, the definition of a reference set remains a requirement (Rousseau, in press). In my opinion, this limitation makes the analyst reflexively aware that each set is a sample and that impact values are sample-dependent.

6. *I3* values correlate both with the number of publications and with the numbers of citations because they are based on the (scalar) sum of the multiplications of these two numbers. Citations themselves can be considered as impact indicators, and publications as performance indicators; they may correlate because of scale effects. In the cases that we tested, the correlations between *I3* and total citations or total publications were higher than the correlations between these latter two (Table 3).

| Indicator | IF-2009 | I3 | Number of publications | Total citations |
|---|---|---|---|---|
| IF-2009 |  | .798 ** | .479 ** | .840 ** |
| I3 | .590 ** |  | .829 ** | .986 ** |
| N of publications | .492 ** | .953 ** |  | .772 ** |
| Total citations | .841 ** | .922 ** | .839 ** |  |

Note: ** Correlation is significant at the 0.01 level (2-tailed); * Correlation is significant at the 0.05 level (2-tailed).

**Table 3**: Rank-order correlations (Spearman's $\rho$; upper triangle) and Pearson correlations *r* (lower triangle) for the 48 journals in the "multidisciplinary" set of the WoS. (Source: Leydesdorff & Bornmann, 2011b, at p. 2142).

In sum, *I3* provides a measure that is statistically warranted and leaves the user free to select from a number of options, such as the choice of a normative evaluation scheme. One can also test heterogeneous sets, such as departments in a university or projects within a program, against one another. The problems with the statistics involved in measuring impact can thus be solved. *I3* can be used both for whole-number counted and fractionally-counted citation rates.

## 5. Next steps

Hitherto, we have not combined the two proposals, but studied *I3* in journals belonging to a single WoS Category (Leydesdorff & Bornmann, 2011b) or specified subsets thereof (Leydesdorff, 2011). If the unit of analysis for an evaluation, however, is multi-disciplinary such as in the case of a university, one can combine the two normalizations and use *I3*-values based on fractional citation scores.

At the level of the WoS or Scopus databases—which are multi-disciplinary in nature—the fractionalization of the citation counts would take care of the differences in "citation potentials" (Garfield, 1979) both synchronically as diachronically (Althouse *et al.*, 2009; Raddicchi & Castellano, 2012, at p. 129) without imposing *a priori* categorization of journals in subject categories. In an email communication (23 June 2010), Ludo Waltman suggested that a remaining difference among fields of sciences might be caused by the different rates at which papers in the last two years are cited in different fields. Correction for this effect would require one additional normalization at the level of each *journal*.

The further introduction of non-parametric statistics into the system may take more time because of existing institutional routines. Most recently, however, both the SCImago Institutions Rankings (at http://www.scimagoir.com/pdf/sir_2011_world_report.pdf) and the Leiden Ranking 2011/2012 (at http://www.leidenranking.com/ranking.aspx) introduced the 10% most-highly cited papers as the *Excellence Index* and *Proportion top-10% publications* ($PP_{top\ 10\%}$), respectively. Bornmann & Leydesdorff (2011) used this same standard for overlays on Google Maps. These excellence indicators for the Scopus and WoS databases, respectively, allow for statistical testing of the significance of differences and rankings (Bornmann *et al.*, in press; Leydesdorff & Bornmann, in press). As noted, the top-10% most-highly cited can be considered as a special case of *I3* (Rousseau, in press). This measure thus has all the advantages listed above. It may be easier to understand this measure than *I3* or its equivalent using six percentile ranks (*PR6*).

Given the increasing consensus about the proportion of the top-10% most-highly cited papers as an excellence indicator, one could also explore this measure as an alternative to the IF. Using fractional counting of the citations and with proper normalization for different document types, the differences of "citation potentials" of journals in different fields of science can significantly be reduced (Leydesdorff & Bornmann, 2011a). The indicator is intuitively simple, allows for statistical testing, and accords with the current state of the art.

Further research is needed because the proportion of 10% most-highly cited documents may insufficiently distinguish among a potentially large group of journals with no or few publications in the top-10%. In the case of patent evaluation, Leydesdorff & Bornmann (in preparation), for example, used the top-25% for this reason (cf. Adams *et al.*, 2000). The top-quartile may be more useful than the top-10% in the case of journals, but this issue has to be informed by empirical research. In addition to the excellence indicator, *I3* and/or *PR6* provide impact indicators which allow for comparisons among less excellent units of analysis by taking also their productivity into account.

**Acknowledgement**

I am grateful to Lutz Bornmann for comments on a previous draft.